\documentclass[preprint,number]{elsarticle}
\usepackage{graphics}
\usepackage{amssymb}
\journal{Physics Letters B}
\makeatletter
\newdimen\z@ \z@=0pt % can be used both for 0pt and 0
\newskip\z@skip \z@skip=0pt plus0pt minus0pt
\def\m@th{\mathsurround=\z@}
\def\ialign{\everycr{}\tabskip\z@skip\halign} % initialized \halign
\def\eqalign#1{\null\,\vcenter{\openup\jot\m@th
  \ialign{\strut\hfil$\displaystyle{##}$&$\displaystyle{{}##}$\hfil
      \crcr#1\crcr}}\,}
\makeatother
\newcommand{\affuni}[2]{Dipartimento di Fisica dell'Universit\`a #1, #2,
  Italy.} 
\newcommand{\affinfn}[2]{INFN Sezione di #1, #2, Italy.}

\newcommand{\affinfnn}[2]{INFN Sezione di #1, #2, Italy.}

\begin{document}
\begin{frontmatter}
\title{Study of the $a_0(980)$ meson via the radiative decay
  $\phi\to\eta\pi^0\gamma$ with the KLOE detector \\
  {\large KLOE Collaboration}}
%\collab{KLOE Collaboration}
\author[Na,infnNa]{F.~Ambrosino}
\author[Frascati]{A.~Antonelli}
\author[Frascati]{M.~Antonelli}
\author[Roma2,infnRoma2]{F.~Archilli}
\author[Roma3,infnRoma3]{C.~Bacci}
\author[Karlsruhe]{P.~Beltrame}
\author[Frascati]{G.~Bencivenni}
\author[Frascati]{S.~Bertolucci}
\author[Roma1,infnRoma1]{C.~Bini}
\author[Frascati]{C.~Bloise}
\author[Roma3,infnRoma3]{S.~Bocchetta}
\author[Frascati]{F.~Bossi}
\author[infnRoma3]{P.~Branchini}
\author[Frascati]{P.~Campana}
\author[Frascati]{G.~Capon}
\author[Frascati]{T.~Capussela}
\author[Roma3,infnRoma3]{F.~Ceradini}
\author[Frascati]{S.~Chi}
\author[Na,infnNa]{G.~Chiefari}
\author[Frascati]{P.~Ciambrone}
\author[Roma1]{F.~Crucianelli}
\author[Frascati]{E.~De~Lucia}
\author[Roma1,infnRoma1]{A.~De~Santis}
\author[Frascati]{P.~De~Simone}
\author[Roma1,infnRoma1]{G.~De~Zorzi}
\author[Karlsruhe]{A.~Denig}
\author[Roma1,infnRoma1]{A.~Di~Domenico}
\author[infnNa]{C.~Di~Donato}
\author[Roma3,infnRoma3]{B.~Di~Micco}
\author[infnNa]{A.~Doria}
\author[Frascati]{M.~Dreucci}
\author[Frascati]{G.~Felici}
\author[Frascati]{A.~Ferrari}
\author[Frascati]{M.~L.~Ferrer}
\author[Roma1,infnRoma1]{S.~Fiore}
\author[Frascati]{C.~Forti}
\author[Roma1,infnRoma1]{P.~Franzini}
\author[Frascati]{C.~Gatti}
\author[Roma1,infnRoma1]{P.~Gauzzi\corref{cor}}
\ead{paolo.gauzzi@roma1.infn.it}
\cortext[cor]{Corresponding author, Dipartimento di Fisica, Sapienza
  Universit\`a di Roma, P.le A.Moro, 2 I-00185 Roma, Italy;
  Tel. +390649914266, Fax +39064957697} 
\author[Frascati]{S.~Giovannella}
\author[Lecce,infnLecce]{E.~Gorini}
\author[infnRoma3]{E.~Graziani}
\author[Karlsruhe]{W.~Kluge}
\author[Moscow]{V.~Kulikov}
\author[Roma1,infnRoma1]{F.~Lacava}
\author[Frascati]{G.~Lanfranchi}
\author[Frascati,StonyBrook]{J.~Lee-Franzini}
\author[Karlsruhe]{D.~Leone}
\author[Frascati,Energ]{M.~Martini}
\author[Na,infnNa]{P.~Massarotti}
\author[Frascati]{W.~Mei}
\author[Na,infnNa]{S.~Meola}
\author[Frascati]{S.~Miscetti}
\author[Frascati]{M.~Moulson}
\author[Frascati]{S.~M\"uller}
\author[Frascati]{F.~Murtas}
\author[Na,infnNa]{M.~Napolitano}
\author[Roma3,infnRoma3]{F.~Nguyen}
\author[Frascati]{M.~Palutan}
\author[infnRoma1]{E.~Pasqualucci}
\author[infnRoma3]{A.~Passeri}
\author[Frascati,Energ]{V.~Patera}
\author[Na,infnNa]{F.~Perfetto}
\author[infnLecce]{M.~Primavera}
\author[Frascati]{P.~Santangelo}
\author[Na,infnNa]{G.~Saracino}
\author[Frascati]{B.~Sciascia}
\author[Frascati,Energ]{A.~Sciubba}
\author[Frascati]{A.~Sibidanov}
\author[Frascati]{T.~Spadaro}
\author[Roma1,infnRoma1]{M.~Testa}
\author[infnRoma3]{L.~Tortora}
\author[infnRoma1]{P.~Valente}
\author[Frascati]{G.~Venanzoni}
\author[Frascati,Energ]{R.Versaci}
\author[Roma1,infnRoma1]{R.~Volpe}
\author[Frascati,Beijing]{G.~Xu}
%%%%%
\address[Frascati]{Laboratori Nazionali di Frascati dell'INFN, 
Via E.Fermi 40, I-00044 Frascati, Italy.}
\address[Karlsruhe]{Institut f\"ur Experimentelle Kernphysik, 
Universit\"at Karlsruhe D-76128 Karlsruhe, Germany.}
\address[Lecce]{\affuni{del Salento}{Via Arnesano, I-73100 Lecce}}
\address[infnLecce]{\affinfn{Lecce}{Via Arnesano, I-73100 Lecce}}
\address[Na]{Dipartimento di Scienze Fisiche dell'Universit\`a 
``Federico II'', Via Cintia, I-80126 Napoli, Italy}
\address[infnNa]{INFN Sezione di Napoli, Via Cintia, I-80126 Napoli, Italy}
\address[Energ]{Dipartimento di Energetica, Sapienza Universit\`a di
  Roma, P.le A.Moro, 2 I-00185 Roma, Italy.}
\address[Roma1]{Dipartimento di Fisica, Sapienza Universit\`a di
  Roma, P.le A.Moro, 2 I-00185 Roma, Italy.}
\address[infnRoma1]{\affinfn{Roma}{P.le A.Moro, 2 I-00185 Roma}}
\address[Roma2]{\affuni{di Roma ``Tor Vergata''}{Via della Ricerca Scientifica, 1
    I-00133 Roma}}
\address[infnRoma2]{\affinfnn{Roma Tor Vergata}{Via della Ricerca
    Scientifica, 1 I-00133 Roma}}
\address[Roma3]{\affuni{di Roma ``Roma Tre''}{Via della Vasca Navale, 84
    I-00146 Roma}}
\address[infnRoma3]{\affinfn{Roma Tre}{Via della Vasca Navale, 84 I-00146
    Roma}}
\address[StonyBrook]{Physics Department, State University of New 
York at Stony Brook, Stony Brook, NY 11794-3840 USA.}
\address[Beijing]{Institute of High Energy 
Physics of Academica Sinica, PO Box 918 Beijing 100049, PR China.}
\address[Moscow]{Institute for Theoretical 
and Experimental Physics, B. Cheremushkinskaya ul. 25, RU-117218 Moscow,
Russia.}
\begin{abstract}
  We have studied the $\phi\to a_0(980)\gamma$ process with the KLOE detector
  at the Frascati $\phi-$factory DA$\Phi$NE by detecting the
  $\phi\to\eta\pi^0\gamma$ decays in the final states with  
  $\eta\to\gamma\gamma$ and $\eta\to\pi^+\pi^-\pi^0$.
  We have measured the branching ratios for both final states:  
  $Br(\phi\to\eta\pi^0\gamma)=(7.01\pm 0.10\pm 0.20)\times 10^{-5}$ and 
  $(7.12\pm 0.13\pm 0.22)\times 10^{-5}$ respectively.
  We have also extracted the $a_0(980)$ mass and its couplings to
  $\eta\pi^0$, $K^+K^-$, and to the $\phi$ meson from the fit of the
  $\eta\pi^0$ invariant mass distributions using different phenomenological
  models. 
\end{abstract}
\end{frontmatter}
\section{Introduction}
The problem of the internal structure of the scalar mesons with mass below 1
GeV is still open\cite{klempt:2007}.
It is controversial whether they are $q\bar q$
mesons\cite{scadron:2004}, $qq\bar q\bar q$
states\cite{jaffe:1977}, bound states of a $K\bar K$
pair\cite{weinstein:1982} or a mixing of these configurations. \\ 
An important part of the program of the KLOE experiment, carried out at the
Frascati $\phi$-factory DA$\Phi$NE, has been dedicated
to  the study of the radiative decays $\phi(1020)\to P_1 P_2\gamma$
($P_{1,2}=$ pseudoscalar mesons).   
These decays are dominated by the exchange of a scalar meson $S$ in
the intermediate state ($\phi\to S\gamma$, and $S\to P_1 P_2$), and both
their branching ratios and the $P_1 P_2$ invariant mass shapes depend on
the scalar structure.\\ 
The $\phi\to\eta\pi^0\gamma$ decay has been already used by KLOE
and by other experiments to study the neutral component of the isotriplet
$a_0(980)$\cite{aloisio:2002,achasov:2000}.
This process is well suited to study the $\phi\to a_0(980)\gamma$ dynamics,
since it is dominated by the scalar production, with small vector
background, contrary to $\pi^0\pi^0\gamma$ and $\pi^+\pi^-\gamma$
cases, where a large irreducible background interferes with the
$f_0(980)$ signal\cite{ambrosino:2007}.\\ 
In this paper the result of the analysis of the $\phi\to\eta\pi^0\gamma$
decay, performed on a sample with 20 times larger statistics than the
previously published paper\cite{aloisio:2002}, is presented.  
The final states corresponding to $\eta\to\gamma\gamma$ and
$\eta\to\pi^+\pi^-\pi^0$ have been
selected.
The $\eta\pi^0$ invariant mass distributions have been fit to two 
models of parametrization of the $\phi\to a_0(980)\gamma$ decay, to extract
the relevant $a_0(980)$ parameters (mass and couplings).
\section{DA$\Phi$NE and KLOE}
The Frascati $\phi$-factory DA$\Phi$NE is an $e^+e^-$ collider operating at
a center of mass energy $\sqrt{s}=M_{\phi}\simeq$ 1020 MeV.
The beams collide at an angle of ($\pi$ - 0.025) rad, thus producing $\phi$
mesons with small momentum ($p_{\phi}\simeq$ 13 MeV) in the horizontal
plane.
The KLOE detector\cite{adinolfi:2002} consists of two main subdetectors: a
large volume drift chamber (DC) and a fine sampling lead-scintillating
fibers electromagnetic calorimeter (EMC).  
The whole apparatus is inserted in a 0.52 T axial magnetic field,
produced by a superconducting coil.
The DC is 3.3 m long, with inner and outer radii of 25 and 200 cm
respectively.
It contains {$12~582$} drift cells arranged in 58 stereo layers uniformly
distributed in the sensitive volume and it is filled with a gas mixture of
90\% helium and 10\% isobutane. 
Its spatial resolution is 200 $\mu$m and the tracks coming from the beam
interaction point (IP) are reconstructed with
$\sigma(p_{\perp})/p_{\perp}\leq 0.4\%$.
The position resolution for two track vertices is about 3 mm.\\
The DC is surrounded by the EMC, that covers 98\% of the solid angle, and
is divided into a barrel, made of 24 trapezoidal modules about 4 m long,
with the fibres running parallel to the barrel axis, and two endcaps of 32
module each, with fibers aligned vertically. 
The read-out granularity is $\sim 4.4\times 4.4$ cm$^2$, for a total of 2440
cells, read at both ends by photomultipliers.
The coordinate of a particle along the fiber direction is reconstructed from
the difference of the arrival time of the signals at the two ends of the
cell. 
Cells close in time and space are grouped together into clusters.
The cluster energy is the sum of the cell energies, while the cluster time
and position are energy weighed averages.
The energy and time resolutions for photons are
$\sigma_E/E=5.7\%/\sqrt{E({\rm GeV})}$ and $\sigma_t=57~{\rm
  ps}/\sqrt{E({\rm GeV})}\oplus 100~{\rm ps}$ respectively.
Cluster positions are measured with resolutions of 1.3 cm in the
coordinates transverse to the fibers, and $1.2~{\rm cm}/\sqrt{E({\rm
    GeV})}$ in the longitudinal coordinate. 
The detection efficiency for photons of $E\simeq 20$ MeV is greater than 80\%
and reaches almost 100\% at $E > 80$ MeV. \\  
The KLOE trigger is based on the detection of two energy
deposits in the EMC, with $E > 50$ MeV in the barrel and  $E > 150$ MeV in
the endcaps. 
\section{Event selection}
The results are based on the data collected during the 2001-02 run, at
$\sqrt{s}\simeq M_{\phi}$.   
Of the two selected decay chains, the fully neutral one is characterized by
high statistics and large background, while the charged one has small
background but lower statistics.
These two decay chains have been selected with different criteria and
slightly different data samples have been used: 414 pb$^{-1}$ for the fully
neutral and 383 pb$^{-1}$ for the charged decay.  
Monte Carlo (MC) samples of signal and of background processes have been
produced with the simulation program of the experiment\cite{ambrosino:2004}.
They have been generated on a run-by-run basis, simulating the machine
operating conditions and background levels as measured in the data.
Three MC samples, generated with different luminosity scale
factors (LSF = $L_{MC}/L_{data}$), have been used: 
\begin{enumerate}
\item the {\it rad} sample contains all the radiative $\phi-$decays plus
  the non resonant process $e^+ e^-\to\omega\pi^0$, with LSF=5;
\item the {\it kk} sample contains $\phi\to K^0\overline{K^0}$ with all
    subsequent kaon decays generated with LSF=1;
\item the {\it all} sample contains all the $\phi$ decays with LSF=1/5; it
  is  used to find possible backgrounds not included in the two main samples.
\end{enumerate} 
The shape of the $\eta\pi^0$ invariant mass distribution has been simulated
according to the spectrum obtained from the previously published
analysis\cite{aloisio:2002}. 
\subsection{$\phi\to\eta\pi^0\gamma$ with $\eta\to\gamma\gamma$}
\begin{figure}[htb]
\begin{tabular}{cc}
  \includegraphics[width=0.45\textwidth]{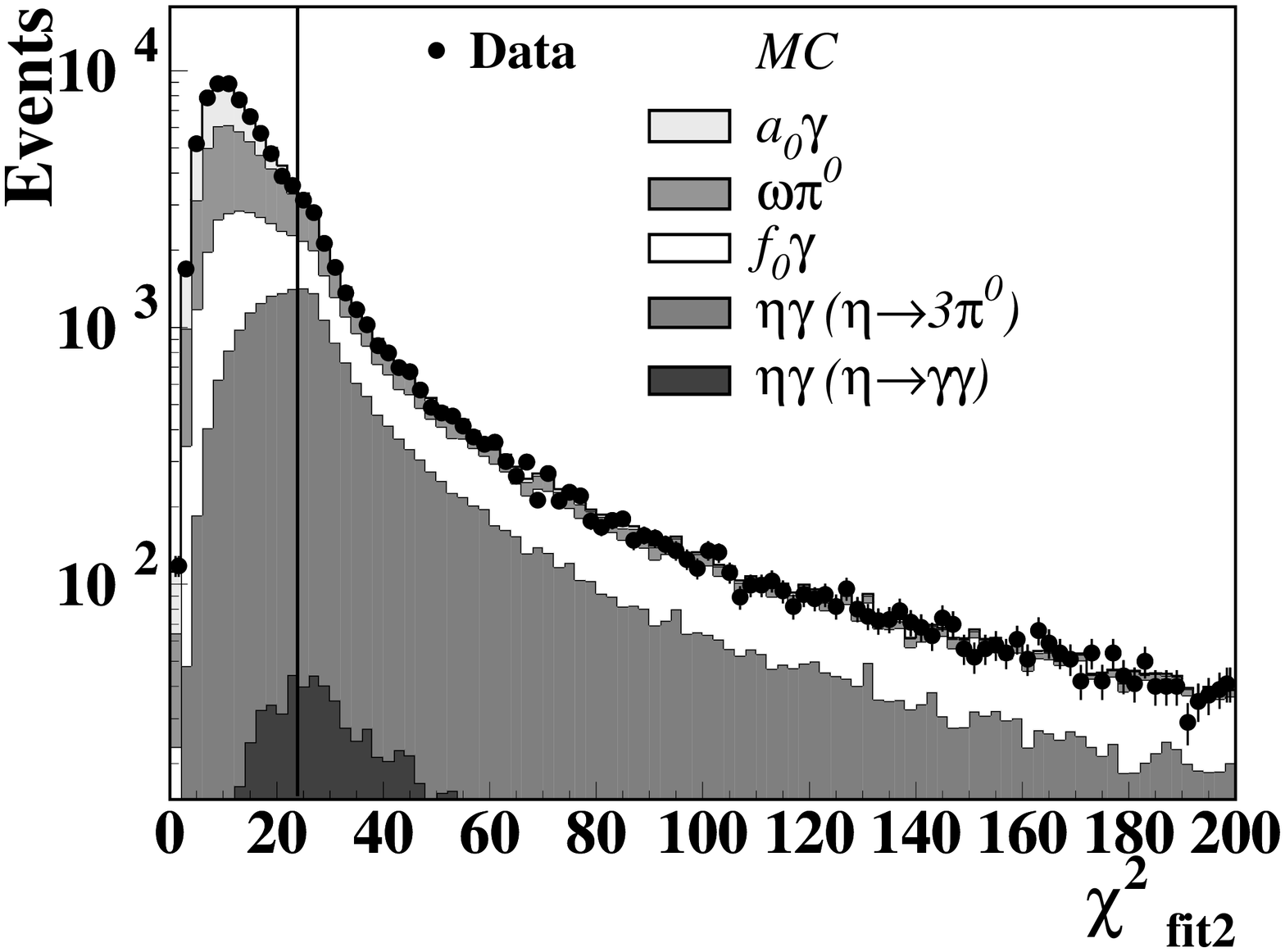} &
  \includegraphics[width=.45\textwidth]{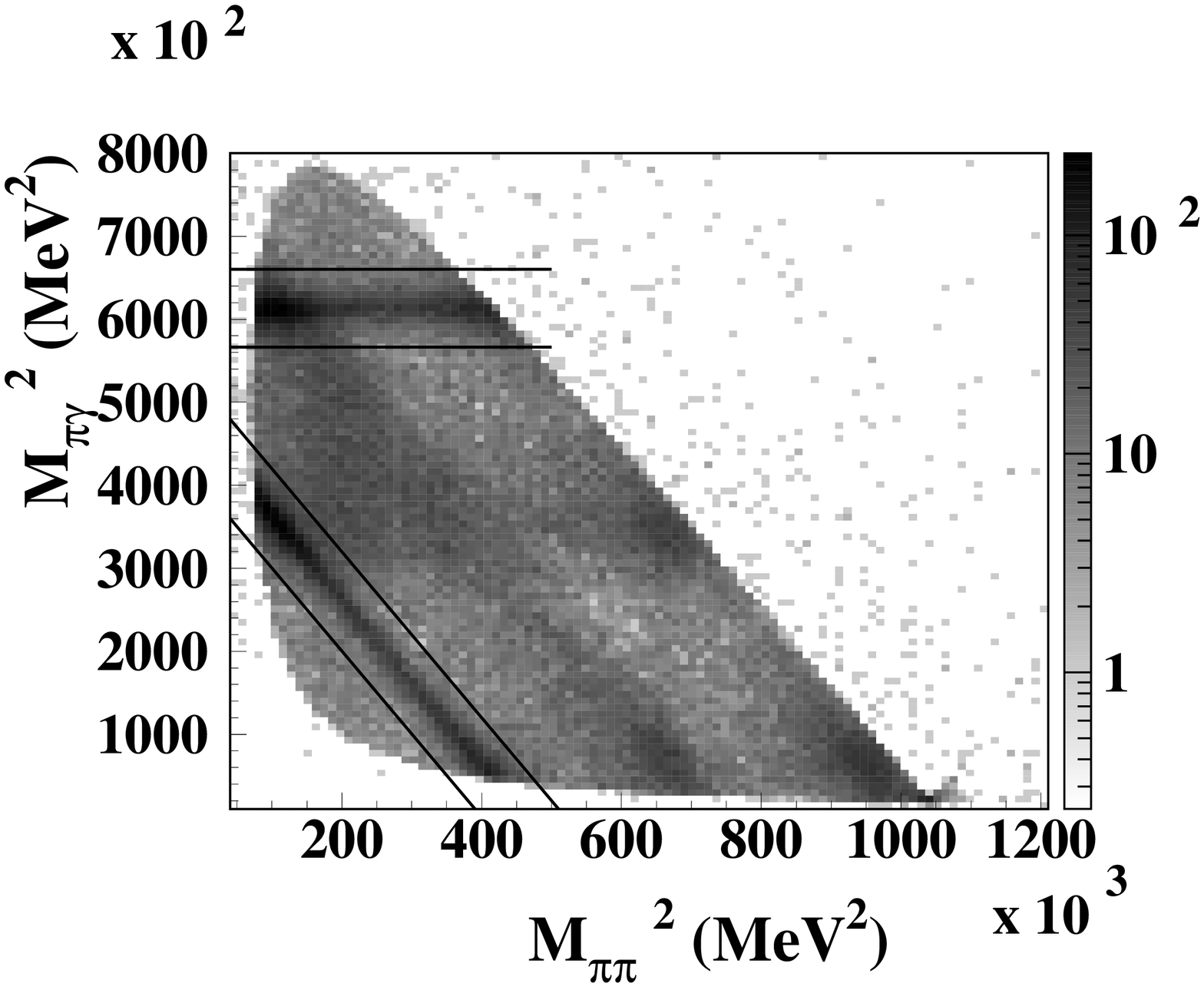} \\ 
\end{tabular}
  \caption{ Right: $\chi^2$ of the second kinematic fit; the
      applied cut at $\chi^2_{fit2} = 24$ is also shown.
 Left: Dalitz plot of data in the background hypothesis ($\pi^0\pi^0\gamma$).}
\label{fig:chi2fit2}
\end{figure}

\begin{figure}[htb]
\begin{tabular}{cc}
  \includegraphics[width=0.45\textwidth]{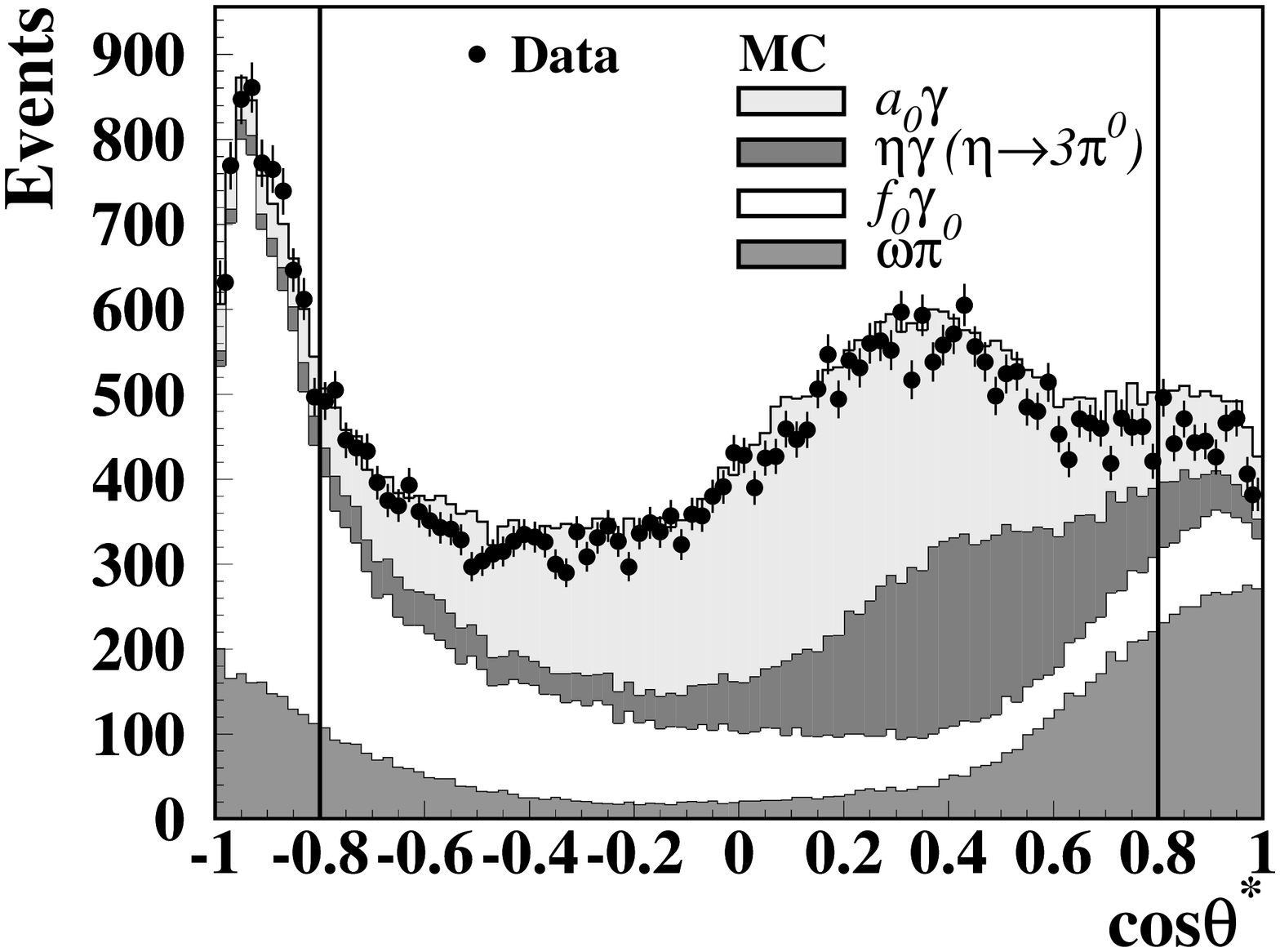} &
  \includegraphics[width=0.45\textwidth]{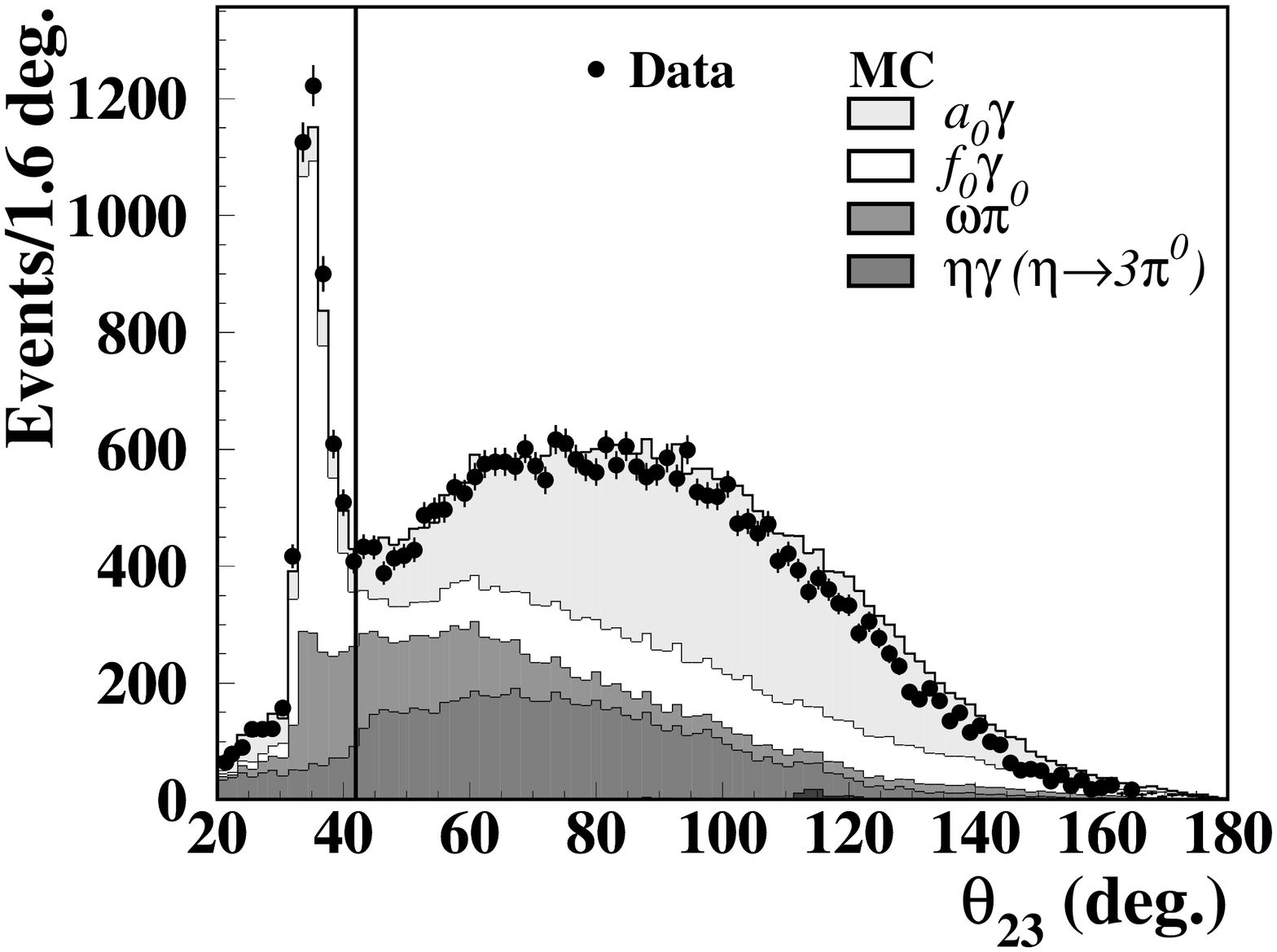}  \\
\end{tabular}
  \caption{ Left: $\cos\theta^{\star}$ distribution
      (see text for explanation); right: angle between the second and
      third photons ordered by decreasing energy (vertical lines represent
      the applied cuts).}
  \label{fig:f0cut}
\end{figure}

\begin{figure}[htb]
\begin{tabular}{cc}
  \includegraphics[width=0.45\textwidth]{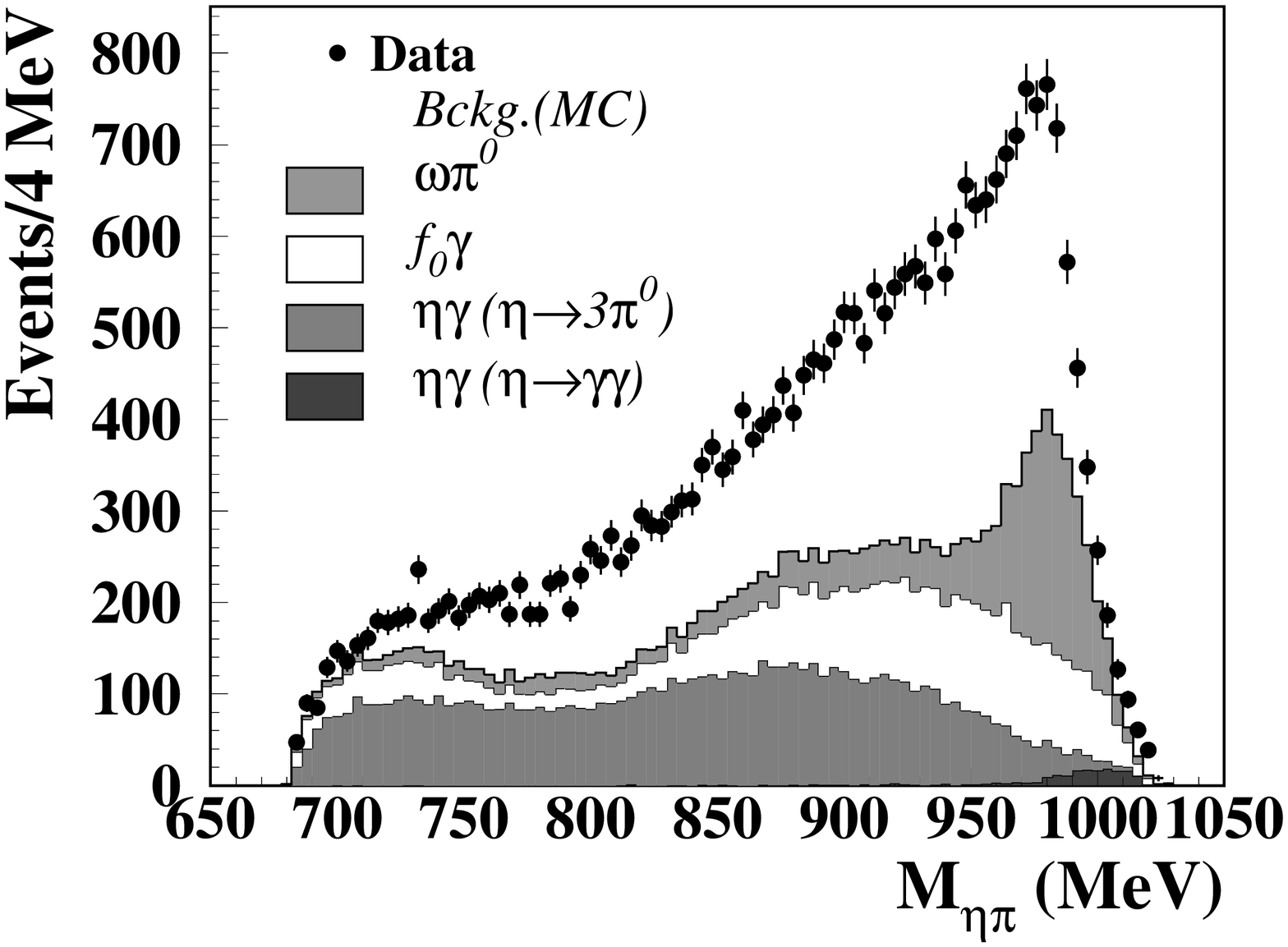} &
  \includegraphics[width=0.45\textwidth]{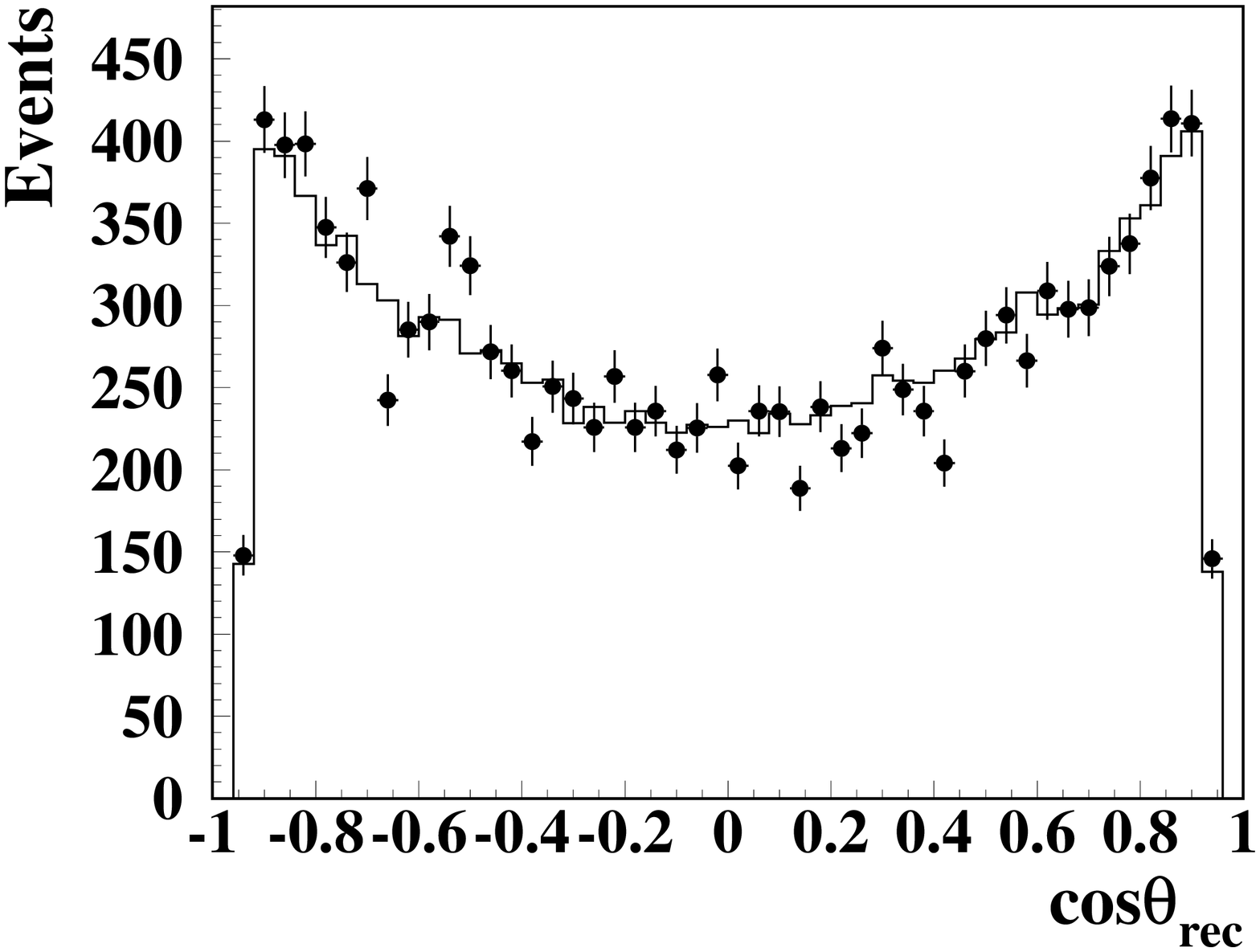} \\
\end{tabular}
  \caption{ Left: $\eta\pi^0$ invariant mass distribution of the
  neutral channel.
  Right: Distribution of the cosine of the polar angle of
      the recoil photon after background subtraction (dots), compared with
  the MC expectation (solid line).}
\label{fig:spectrum}
\end{figure}

This final state is characterized by five prompt photons originating from
the IP.
A prompt photon is defined as an EMC cluster not associated to any charged
track in the DC and satisfying the condition $|t-r/c| < {\rm
  min}[5\sigma_t(E), 2~{\rm ns}]$, where $t$ is the photon flight time, $r$
is the corresponding path length, and $c$ is the speed of light.   
Events with exactly five prompt clusters, with E $>$ 3 MeV and polar
angle $\vartheta > 21^{\circ}$ with respect to the beam line, are selected.\\  
The main background originates from the other five photon final states,
$\phi\to f_0(980)\gamma\to\pi^0\pi^0\gamma$ and
$e^+e^-\to\omega\pi^0\to\pi^0\pi^0\gamma$, and from the seven photon
process, $\phi\to\eta\gamma$ with $\eta\to 3\pi^0$, which can mimic five
photon events due to either photon loss or cluster merging.
Also the three photon final states, $\phi\to\eta(\pi^0)\gamma$ with
$\eta(\pi^0)\to\gamma\gamma$, give a small contribution to the selected
sample, when fake clusters are produced either by accidental coincidence
with machine background or by cluster splittings. 
Other background processes are negligible. \\
The following analysis steps are then applied to the selected events.
\begin{enumerate}
\item First kinematic fit which imposes the total 4-momentum
   conservation and the speed of light for each photon, with 9
   degrees of freedom. 
   Events with $\chi^2_{fit1} > 27$ are rejected.
   A cut at 980 MeV on the total energy of the three most
   energetic photons is also applied to reject residual three photon events
   (processes 4 and 5 of Table \ref{tab:bckg}).
\item Search for the best photon pairing to $\eta$'s and $\pi^0$'s,  by
  choosing the combination that minimizes the $\chi^2$-like variable
  ($i,j,k,l=1,...,5$ are the photon indices):
  \begin{displaymath}
    \chi^2_{pair}=\frac{(M_{ij}-M_{P_1})^2}{\sigma^2_{M_{P_1}}}
    +\frac{(M_{kl}-M_{P_2})^2}{\sigma^2_{M_{P_2}}} 
  \end{displaymath}
  for both $P_1 P_2 = \eta\pi^0$ (signal) or $\pi^0\pi^0$ (background)
  hypotheses. 
$\sigma_{M_{\pi^0}}$ and $\sigma_{M_{\eta}}$ are the width of the $\pi^0$
  and $\eta$ peaks after the first kinematic fit ($\sigma_{M_{\pi^0}}=6$
  MeV and   $\sigma_{M_{\eta}}=9$ MeV).   
\item Second kinematic fit with the two additional constraints of the
  masses of the intermediate particles.
  The number of degrees of freedom is 11.
\end{enumerate}
Background from process 1 and 3 of Table \ref{tab:bckg} dominates the tail of
the distribution of the $\chi^2_{fit2}$ of the second kinematic fit, as
shown in Fig.\ref{fig:chi2fit2}, and it can be reduced by 
cutting at $\chi^2_{fit2} < 24$. 
By using the photon pairing in the background hypothesis,
$\pi^0\pi^0\gamma$, the Dalitz plot of Fig.\ref{fig:chi2fit2} is obtained:
the $f_0\gamma$ background populates the lower right corner, while the two
straight bands are the contribution of $\omega\pi^0$. 
The $a_0$ signal is contained in the region between these bands.
The $\omega\pi^0$ background is strongly reduced by cutting out the two
bands shown in Fig.\ref{fig:chi2fit2}.\\
Assuming the background hypotesis $\omega\pi^0$, the angle $\theta^{\star}$
between the non associated photon and the $\omega$ flight direction can be
defined.   
The regions at large $|\cos\theta^{\star}|$ (Fig.\ref{fig:f0cut}.left)
are dominated by $\omega\pi^0$ and $f_0\gamma$ backgrounds.
The cut $|\cos\theta^{\star}| < 0.8$ is then applied.
Another effective cut to reduce the $f_0\gamma$ background is $\theta_{23}
> 42^{\circ}$ (Fig.\ref{fig:f0cut}.right), where $\theta_{23}$ is the angle
between the second and third photons ordered by decreasing energy.    \\
After these cuts the overall selection efficiency, evaluated by MC, is
almost independent from the $\eta\pi^0$ invariant mass and its average value
is 38.5\%. 
The final sample consists of {29 601} events and the expected S/B ratio is
about 1.0 (see Table \ref{tab:bckg}).   
The residual background is irreducible and has to be evaluated and
subtracted. 
A reweighing procedure has been adopted: 
\begin{enumerate}
  \item for each specific background process a data sample with a small
    signal content (below few percent) has been selected;
  \item a fit has been performed on selected kinematical distributions,
    using the corresponding MC shapes to determine the weight to be
    assigned to that  specific background; the weight is defined as the
    ratio of the number of events found by the fit to the number of
    expected events from MC.  
\end{enumerate} 
In the last two columns of Table \ref{tab:bckg} the applied weights and the
numbers of background events in the final sample are listed. 
\begin{table}[htb]
  \caption{Background processes for $\phi\to\eta\pi^0\gamma$, with
    $\eta\to\gamma\gamma$.  
(S/B)$_1$ is the signal to background ratio after the preselection,
(S/B)$_2$ the same ratio at the end of the whole analysis chain.
    The reweighing factors, $w$, are also listed.
  Last column reports the final background estimate.}
\begin{center}
  \begin{tabular}{clccc|c}\hline
 &    Process                                         & (S/B)$_1$ &
    (S/B)$_2$ & $w$ & Background events\\ 
    \hline
    1 &   $\phi\to f_0\gamma\to\pi^0\pi^0\gamma$          & 0.40 & 4.4 &
    1.2 & 5062 $\pm$ 60   \\
    2 &   $e^+e^-\to\omega\pi^0\to\pi^0\pi^0\gamma$       & 0.14 & 3.1 &
    0.96 & 3825 $\pm$ 37 \\
    3 &   $\phi\to\eta\gamma$ with $\eta\to 3\pi^0$       & 0.10 & 2.8 &
    1.1 & 7248 $\pm$ 78 \\
    4 &   $\phi\to\eta\gamma$ with $\eta\to\gamma\gamma$  & 1.6  & 200 &
    2.5 & 197 $\pm$ 11 \\
    5 &   $\phi\to\pi^0\gamma$                            & 10  &  -- &  --
    & -- \\
    \hline
     & Total background                                  & 0.05 & 1.0 &  & {16
    332} $\pm$ 86 \\
    \hline
  \end{tabular}
\end{center}
  \label{tab:bckg}
\end{table}
The uncertainties are the combination of MC statistics and of the
systematics on the applied weights. 
The correlations have also been taken into account.
After the background subtraction the number of signal event is {13 269}
$\pm$ 192.  
In Fig.\ref{fig:spectrum} the $\eta\pi^0$ invariant mass distribution of
the final sample is shown together with the background contributions. 
The invariant mass resolution is about 4 MeV, with non-gaussian tails
mainly due to wrong photon combinations. 
In the same figure, the distribution of the polar angle $\theta_{rec}$ of
the recoil photon is plotted after the background subtraction: good
agreement with the expected $1+\cos^2\theta_{rec}$ behaviour is shown.  
\subsection{$\phi\to\eta\pi^0\gamma$ with $\eta\to\pi^+\pi^-\pi^0$}
\begin{figure}[htb]
\begin{center}
\begin{tabular}{cc}
  \includegraphics[width=0.45\textwidth]{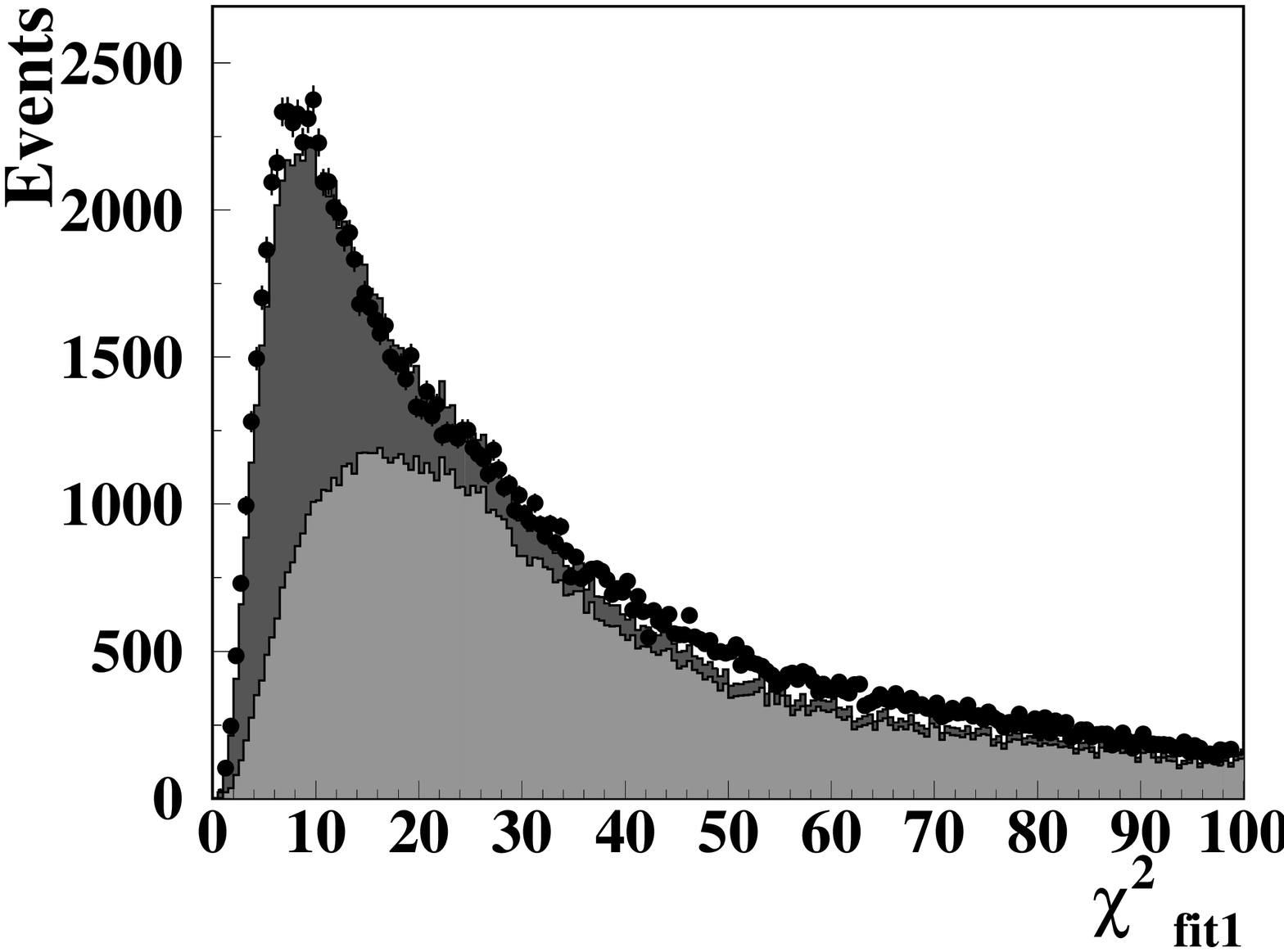} &
  \includegraphics[width=0.45\textwidth]{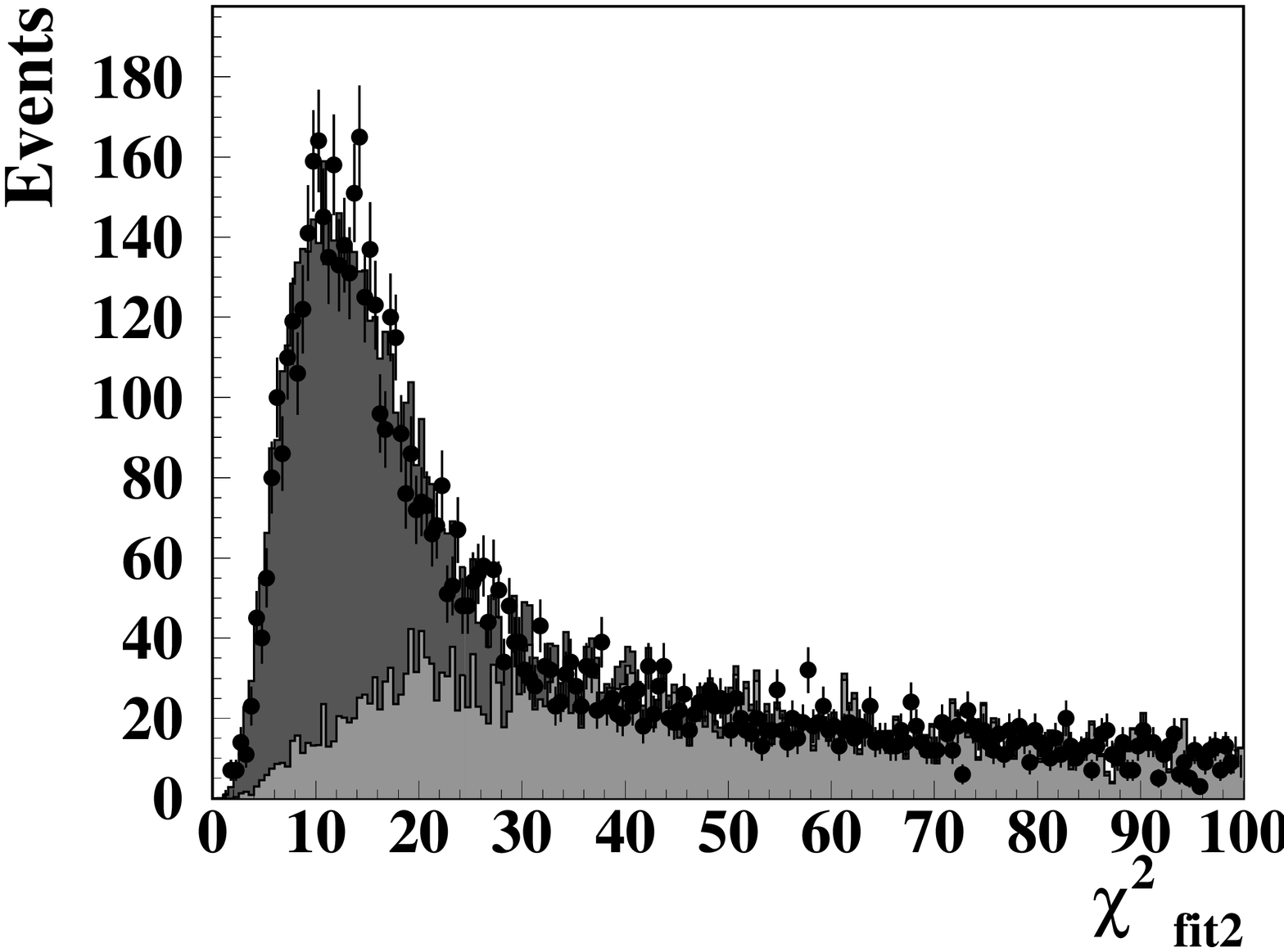} \\
\end{tabular}
  \caption{ $\chi^2$ distributions for the first (left) and
  second (right) kinematic fit. The selected data sample (points) is
  compared to the MC  expectation (dark grey histograms) given by the
  weighed sum of the signal and the estimated background (light grey
  histograms).}  
\label{fig:chisquadri}
\end{center}
\end{figure}

\begin{figure}[htb]
\begin{tabular}{cc}
  \includegraphics[width=0.45\textwidth]{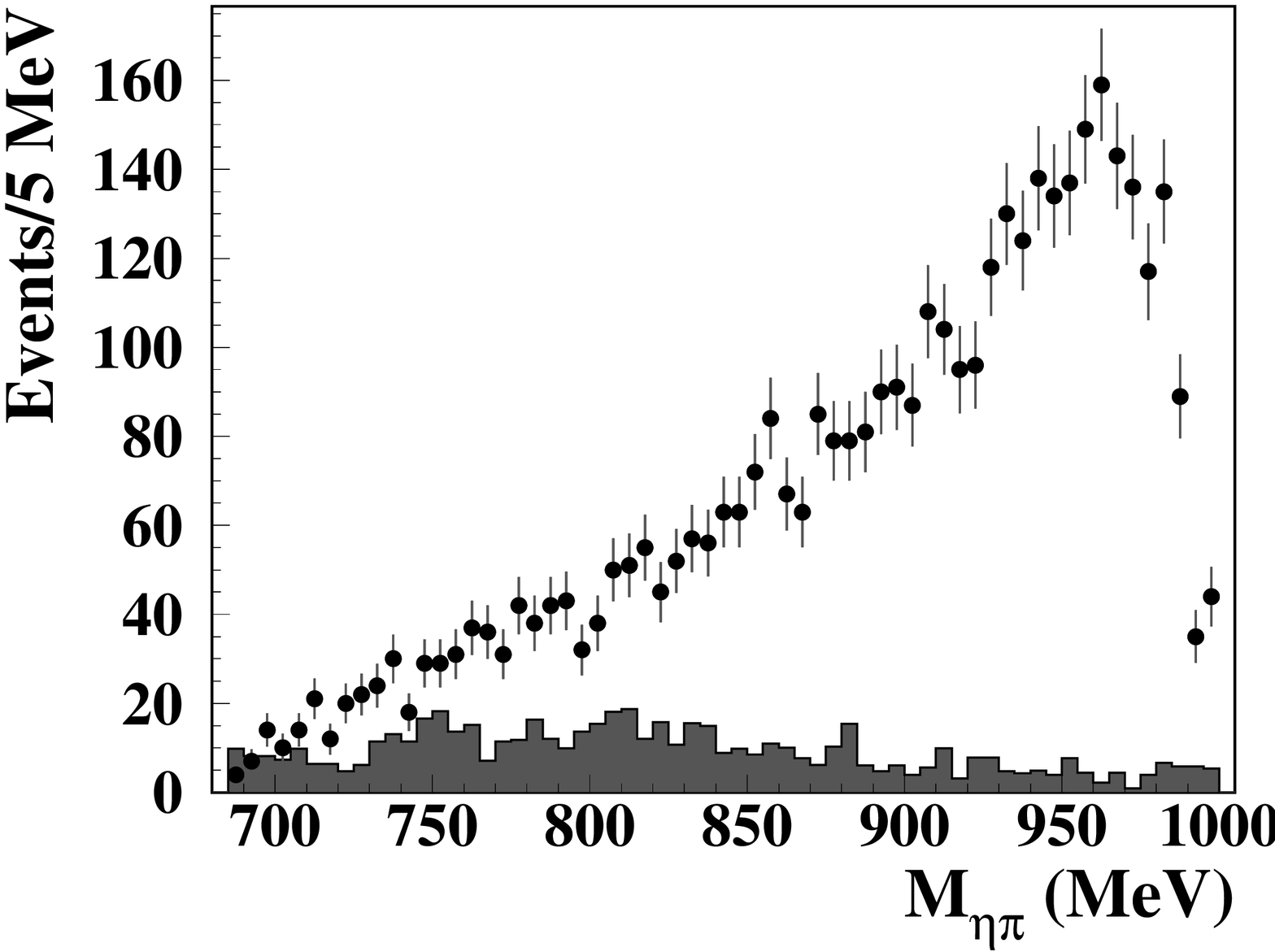} & 
  \includegraphics[width=0.45\textwidth]{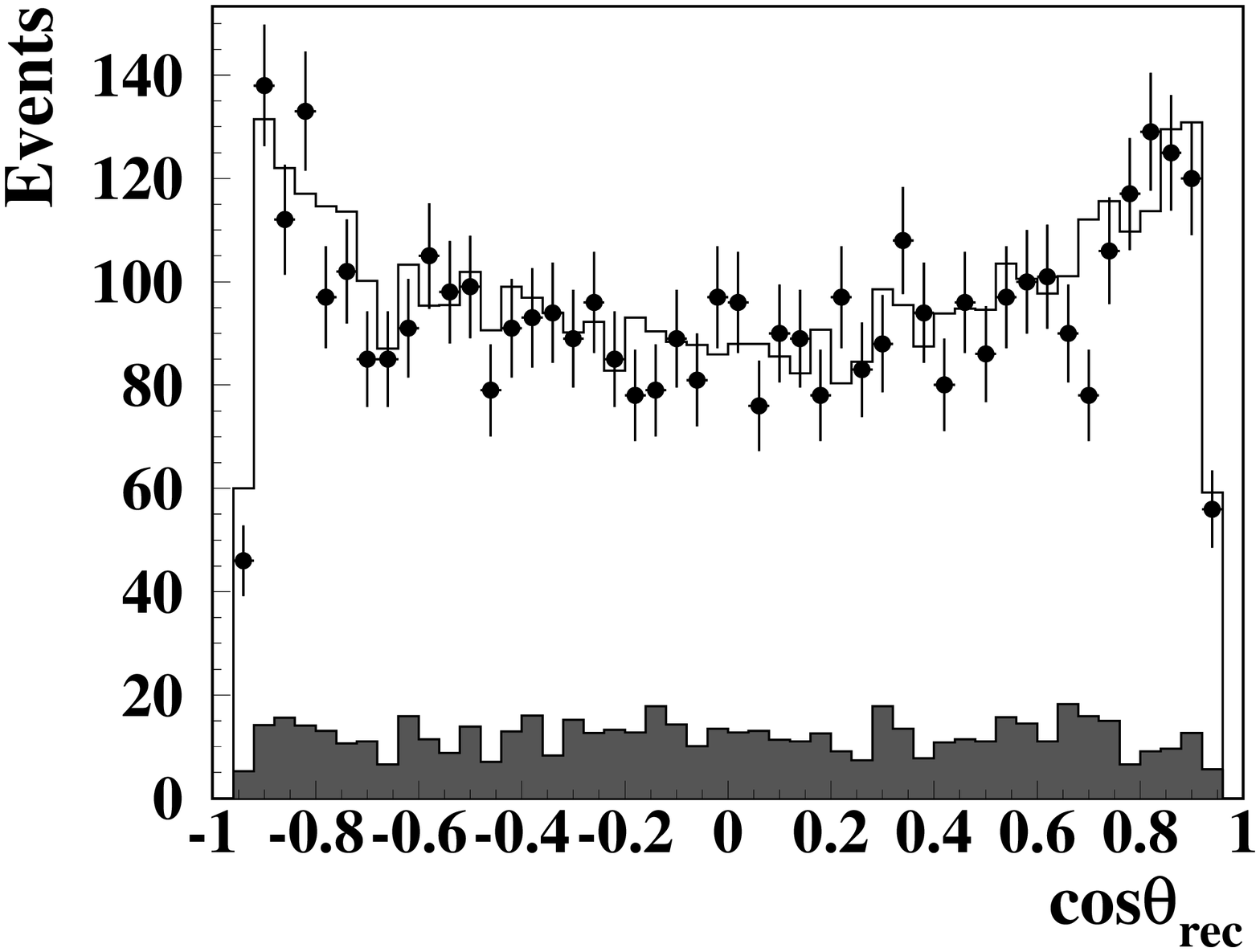} \\
\end{tabular}
  \caption{  Left: $\eta\pi^0$ invariant mass distribution for
      the final data sample (points) compared to the estimated background
      (dark histogram). Right: polar angle of the recoil photon for
      data (points) and for MC expectations (histogram). Dark histogram
      represents the background.}
\label{fig:spectrum_ch}
\end{figure}

With respect to the fully neutral one, this decay provides a lower
statistics since the branching ratio of $\eta\to\pi^+\pi^-\pi^0$ is smaller
than for $\eta\to\gamma\gamma$.  
Moreover a lower acceptance is expected due to the larger number of
particles to be detected.   
However in this case there is a smaller background contamination, since no
other final state with two tracks and five photons has a significant
branching ratio from the $\phi$.   
The main sources of background are due to final states with two tracks and
either four or six photons.  
In order of importance there are: $e^+ e^-\to\omega\pi^0$ with
$\omega\to\pi^+\pi^-\pi^0$ and a fake cluster; $\phi\to K_SK_L$ with
$K_S\to\pi^+\pi^-$ and prompt $K_L\to 3\pi^0$ with one photon lost;
$\phi\to K_SK_L$ with $K_S\to\pi^0\pi^0$ and prompt 
$K_L\to \pi^+\pi^-\pi^0$ or $\pi\ell\nu$ with either one photon lost or one
fake cluster; $\phi\to\eta\gamma$ with $\eta\to\pi^+\pi^-\pi^0$ plus
two fake clusters. \\ 
The signal preselection requires the detection of two charged tracks and of
five photons.
The following requirements are then applied:
\begin{enumerate}
\item a vertex with two opposite sign tracks in a cylinder, around the IP,
  of 5 cm radius and 11 cm length; 
\item five prompt photons with $E>$10 MeV; 
\item total energy in the range $900<E_{tot}<1160$ MeV and total momentum
  $|\vec{P_{tot}}|<110$ MeV/c; 
\item{the scalar sum of the momenta of the two pions
    $P_{\Sigma}=|\vec{p_1}|+|\vec{p_2}|$, outside the range
    $418<P_{\Sigma}<430$ MeV/c, which identifies events with
    $K_S\to\pi^+\pi^-$.} 
\end{enumerate}
Events surviving this preselection go to the kinematic fit stage,
similar to that of the neutral channel.
\begin{enumerate} 
\item A kinematic fit with 9 degrees of freedom is performed by imposing
  only the total 4-momentum conservation and speed of light for the
  photons; events with $\chi^2_{fit1}<17$ are retained. 
\item Photons are combined to build $\pi^0$'s and
  $\eta$'s.
There are 15 possibilities to get two $\pi^0$'s out of five photons. 
For each of them there are two choices in the association of one $\pi^0$ to
the $\pi^+\pi^-$ pair. 
For each of these 30 combinations $\chi_{pair}^2$ is computed according to
($i,j,k,l=1,...,5$ are the photon indices):
\begin{displaymath}
  \chi^2_{pair}={{(M_{ij}-M_{\pi^0})^2}\over {\sigma^2_{M_{\pi^0}}}}+
  {{(M_{kl}-M_{\pi^0})^2}\over {\sigma^2_{M_{\pi^0}}}}+
  {{(M_{\pi^+\pi^-\pi^0}-M_{\eta})^2}\over {\sigma^2_{M_{\eta}}}}
\end{displaymath}  
Events with at least one combination with $\chi_{pair}^2<$ 10 are retained.  
\item The second kinematic fit is performed on all the combinations selected
  by the previous step adding the three mass constraints, for a total of 12
  degrees of freedom.
The combination with the lowest $\chi^2_{fit2}$ is chosen. 
Only events with $\chi^2_{fit2}<20$ are retained. 
\item Finally, events with the recoil photon energy below 20 MeV are
  discarded to remove events with a spurious low energy photon.  
\end{enumerate}
The final sample consists of 4181 events. 
The overall selection efficiency for the signal, evaluated by MC, is 19.4\%,
almost independent from the  $\eta\pi^0$ invariant mass, decreasing only
at very high invariant mass values.  
Fig.\ref{fig:chisquadri} shows the data-MC agreement for the $\chi^2$
distributions of the first and second kinematic fits.  
The MC distributions include signal and background events.
The mass resolution is about 4 MeV for all mass values, with non gaussian
tails, mainly due to events with a wrong photon combination.\\
The residual background is evaluated by applying the selection
procedure on MC samples and by checking the absolute normalization on
background enriched data control samples.
In order to properly normalize the observed numbers of events,
data and MC samples after the preselection but before the kinematic fit
have been used.  
At this level the expected contribution of the signal does not exceed $2
\div 3$\%.  
Four variables have been chosen to compare data and MC samples: $E_{tot}$,
$|\vec{P_{tot}}|$, $M_{\gamma\gamma}$ and $M_{\pi\pi\gamma\gamma}$ where
$M_{\gamma\gamma}$ is the invariant mass of any pair of photons (10
combinations per event) and $M_{\pi\pi\gamma\gamma}$ is the invariant mass
of the two pions and any pair of two photons (again 10 combinations per
event).  
The four distributions for the data are simultaneously fit with the weighed
sum of the same MC distributions for each background sample and for the
signal.  
The weights of the {\it rad} and {\it kk} samples are the free
parameters.
$w_{rad}=0.45$ and $w_{kk}=1.3$ are obtained, from which the numbers of
background events $B_{rad}=307$ and $B_{kk}=264$ are estimated.  
8 additional background events from the $all$ sample have also to be taken
into account.
The fit has been repeated separately on each control distribution and the
spread obtained in the estimated number of events is taken as systematic 
uncertainty. 
The total number of background events is $579\pm 27$, where the
uncertainty is the quadratic sum of the statistical and the systematic
uncertainties.    
This background accounts for about 14\% of the selected events. \\
Fig.\ref{fig:spectrum_ch} shows the $\eta\pi^0$ invariant mass distribution. 
In the same figure, the distribution of the polar angle of the recoil
photon is shown, and is compared to the MC expected behaviour. 
Also in this case the distribution agrees with the $1+\cos^2\theta_{rec}$
dependence of the signal.  
\subsection{Branching ratio evaluation}
The branching ratio of the process $\phi\to\eta\pi^0\gamma$ is
obtained from the formula:
\begin{equation}
  Br(\phi\to\eta\pi^0\gamma)=\frac{N_{f}-B_{f}}{\varepsilon_{f}
      N^{(f)}_{\phi}Br(\eta\to f)} \;\;\;\;\; (f=\gamma\gamma,
      \pi^+\pi^-\pi^0)
\label{eq:br} 
\end{equation}
where $N_{f}$ is the total number of selected events, $B_{f}$ the estimated
background, $\varepsilon_f$ is the average efficiency. 
$N_{\phi}$ is the number of produced $\phi$ mesons evaluated from the
number $N_{\eta\gamma}$ of $\phi\to\eta\gamma$ with
$\eta\to\pi^0\pi^0\pi^0$ events.  
\begin{equation}
  N_{\phi} =
  \frac{N_{\eta\gamma}}{\varepsilon_{\eta\gamma} Br(\phi\to\eta\gamma)
    Br(\eta\to\pi^0\pi^0\pi^0)}
\label{eq:norm}
\end{equation}
The $Br(\pi^0\to\gamma\gamma)$ is not included in eq.(\ref{eq:br}) and
(\ref{eq:norm}) since it has been already taken into account in the MC.
The normalization sample has been selected by requiring no tracks in the DC
and six or more prompt clusters in the EMC, in the same runs used for the
signal selection.  
$N_{\eta\gamma}=4.2\times 10^6$ events have been found in the sample used
for the analysis of the fully neutral decay chain, with efficiency
$\varepsilon_{\eta\gamma} = $ 81\%, corresponding to
$N^{(\gamma\gamma)}_{\phi}=(1.24\pm 0.03)\times 10^9$. \\
By using $Br(\eta\to\gamma\gamma)=(39.31\pm 0.20)\%$\cite{amsler:2008}, the
branching ratio is obtained:
\begin{equation}
  Br(\phi\to\eta\pi^0\gamma)=(7.01\pm 0.10 \pm 0.20)\times 10^{-5} 
\label{eq:brneutral}
\end{equation}
The first uncertainty is due to statistics and to the background
subtraction. 
Several sources of systematics have been taken into account
(see Table \ref{tab:syst}): photon counting (dominated by the detection
efficiency for low energy photons), the data-MC discrepancies in the
evaluation of the selection efficiency, and the normalization uncertainty.\\ 
\begin{table}[htb]
  \caption{Main sources of systematic uncertainty on
     the branching ratio (\ref{eq:brneutral}).}
\begin{center}
  \begin{tabular}{lc}\hline
    Source & uncert. ($\times 10^{-5}$) \\
    \hline
    Photon counting & 0.08 \\
    Selection efficiency & 0.12 \\
    $Br(\eta\to\gamma\gamma)$ & 0.04 \\
    $Br(\phi\to\eta\gamma)$ & 0.13 \\
    $Br(\eta\to\pi^0\pi^0\pi^0)$ & 0.05  \\
    \hline
  \end{tabular}
\end{center}
  \label{tab:syst}
\end{table}
The data sample analyzed for the charged decay channel is slightly smaller
than the other one, $N^{(\pi^+\pi^-\pi^0)}_{\phi}=(1.15\pm 0.03)\times
10^9$.  
By using $Br(\eta\to\pi^+\pi^-\pi^0)=(22.73\pm
0.28)\%$\cite{amsler:2008} 
\begin{equation}
  Br(\phi\to\eta\pi^0\gamma)=(7.12\pm0.13\pm0.22)\times 10^{-5}
\label{eq:brcharged}
\end{equation}
is obtained.
The first uncertainty is the quadratic sum of the statistical uncertainty on
$N_{\pi^+\pi^-\pi^0}$ and of the uncertainty on the background; the second
one is systematic, mainly due to the absolute normalization, and
includes a 1\% error due to the efficiency evaluation.\\ 
The two branching ratios (\ref{eq:brneutral}) and (\ref{eq:brcharged})
are compatible with the old KLOE results: $(8.51 \pm 0.51 \pm 0.57)\times
10^{-5}~(\eta\to\gamma\gamma)$ and $(7.96\pm 0.60 \pm 0.40)\times
10^{-5}~(\eta\to\pi^+\pi^-\pi^0)$\cite{aloisio:2002}.   
By combining the two results, taking into account the common normalization
error
\begin{equation}
  Br(\phi\to\eta\pi^0\gamma)=(7.06 \pm 0.22)\times 10^{-5}
\label{eq:avebr}
\end{equation}
is obtained, where the uncertainty is both statistic and systematic. 
\section{Fit of the $\eta\pi^0$ invariant mass distributions}
\label{sec:fit}
In order to extract the relevant parameters of the $a_0$, a
simultaneous fit, with the same set of free parameters, has 
been performed on the two $\eta\pi^0$ invariant mass distributions, by
minimizing the following $\chi^2$: 
\begin{displaymath}
  \chi^2=\sum_{f=\gamma\gamma,\pi^+\pi^-\pi^0}\sum_{i=1}^{n_f}\frac{(N_i^{(f)}
  -B_i^{(f)}-E_i^{(f)})^2}{{\sigma_i^{(f)}}^2} 
\end{displaymath}
where $n_f$ is the number of bins of respectively the fully
neutral and charged $\eta\pi^0$ mass distribution; $N_i$ is the content of the
$i$-th bin and $B_i$ is the number of background events to be subtracted
from the $i$-th bin.
The expected number of events, $E_i$, can be written as
\begin{displaymath}
  E_i^{(f)}=
  N_{\phi}^{(f)}\sum_{j=1}^{n_{f}}\varepsilon_{ij}^{(f)}
  \frac{1}{\Gamma_{\phi}}\int_{{\rm
  bin~}j}\frac{d\Gamma_{th}(\phi\to\eta\pi^0\gamma)}{dm}dm\times Br(\eta\to
  f) 
\end{displaymath}
where $m=M_{\eta\pi^0}$, and $\Gamma_{\phi}=4.26$ MeV\cite{amsler:2008}.
$\varepsilon_{ij}^{(f)}$ is the efficiency matrix (also referred to as
smearing matrix), representing the probability of a signal event with
``true'' mass in the $j$-th bin of the spectrum to be reconstructed in the
$i$-th bin.   
The efficiency matrices, evaluated by MC, are almost diagonal; the
off-diagonal elements take into account resolution effects as well as
wrong photon pairings. 
The differential decay width $d\Gamma_{th}/dm$ has been parametrized
according to two different models. \\
In the ``Kaon Loop'' (KL) model\cite{achasov:1989} the $\phi$ is coupled to
the scalar meson through a loop of charged kaons.
The theoretical function can be written as:
\begin{equation}
  \frac{d\Gamma_{th}(\phi\to\eta\pi^0\gamma)}{dm}=
  \frac{d\Gamma_{scal}}{dm}+\frac{d\Gamma_{vect}}{dm}
  +\frac{d\Gamma_{interf}}{dm}  
  \label{eq:klformula}
\end{equation} 
The scalar term $d\Gamma_{scal}/dm$ is described in some details in
Appendix \ref{app:kl}.  
$d\Gamma_{vect}/dm$, is dominated by $\phi\to\rho\pi^0$ with
$\rho\to\eta\gamma$ and is described in the framework of the Vector
Dominance Models (VDM)\cite{achasov:2001}.
Last term is the interference between the scalar and the vector
amplitudes. \\ 
The free fit parameters are: the $a_0$ mass, the couplings $g_{a_0K^+K^-}$,
$g_{a_0\eta\pi^0}$, the branching ratio of the vector contribution, the
relative phase $\delta$ between scalar and vector amplitudes, and, as a
relative normalization between the two different final states, the ratio 
$R_{\eta}=Br(\eta\to\gamma\gamma)/Br(\eta\to\pi^+\pi^-\pi^0)$.  \\
An alternative parametrization of the amplitude of the decay 
$\phi\to\eta\pi^0\gamma$ has been also used, following
ref.\cite{isidori:2006}.
A point-like coupling of the scalar to the $\phi$ meson is assumed, hence
this model will be called ``No Structure'' (NS) in the following. 
The scalar meson is parametrized as a Breit-Wigner interfering with a
polynomial scalar background and with a vector background (see Appendix
\ref{app:ns}). 
The free parameters in this case are the couplings $g_{a_0K^+K^-}$,
$g_{a_0\eta\pi^0}$, and $g_{\phi a_0\gamma}$, the ratio $R_{\eta}$, the
branching ratio of the vector background, and two complex coefficients, b$_0$
and b$_1$, of the scalar background.
The $a_0$ mass is fixed to avoid fit instabilities, due the large number of
free parameters, and due to the large cancellations that occur among the
terms of eq.(\ref{eq:nsfunction}). 
The chosen value of the $a_0$ mass is the result of the KL fit. \\
\begin{table}[htb]
  \caption{ Fit results for KL and NS models.}
\begin{center}
  \begin{tabular}{l|c|c}\hline
 & KL & NS \\
\hline
    $M_{a_0}$ (MeV)                & 982.5 $\pm$ 1.6 $\pm$ 1.1  & 982.5 (fixed) \\
    $g_{a_0K^+K^-}$ (GeV)          & 2.15 $\pm$ 0.06 $\pm$ 0.06 & 2.01 $\pm$ 0.07 $\pm$ 0.28 \\
    $g_{a_0\eta\pi^0}$ (GeV)       & 2.82 $\pm$ 0.03 $\pm$ 0.04 & 2.46 $\pm$ 0.08 $\pm$ 0.11 \\
    $g_{\phi a_0\gamma}$ (GeV$^{-1}$) & & 1.83 $\pm$ 0.03 $\pm$ 0.08 \\
    $\delta$ (deg.)                & 222 $\pm$ 13 $\pm$ 3 & \\
    B.r. of vector backg. ($\times 10^6$)  & 0.92 $\pm$ 0.40
    $\pm$ 0.15 & $\sim $ 0 \\
    $R_{\eta}$                 & 1.70 $\pm$ 0.04 $\pm$ 0.03 & 1.70 $\pm$ 0.03 $\pm$ 0.01 \\
    $|{\rm b}_0|$                       &   & 14.9 $\pm$ 0.6 $\pm$ 0.5 \\ 
    $arg({\rm b}_0)$ (deg.)          &   & 38.3 $\pm$ 1.1 $\pm$ 0.6 \\
    $|{\rm b}_1|$                      &   & 21.3 $\pm$ 1.4 $\pm$ 0.9 \\ 
    $arg({\rm b}_1)$ (deg.)          &   & 57.3 $\pm$ 1.4 $\pm$ 1.1 \\
    $\chi^2/ndf$            & 157.1 / 136        & 140.6 / 133        \\
    $P(\chi^2)$                & 10.4\%            & 30.9\%            \\
    \hline
  \end{tabular}
\end{center}
  \label{tab:fitkl}
\end{table}
\begin{figure}[htb]
\begin{tabular}{cc}
  \includegraphics[width=0.45\textwidth]{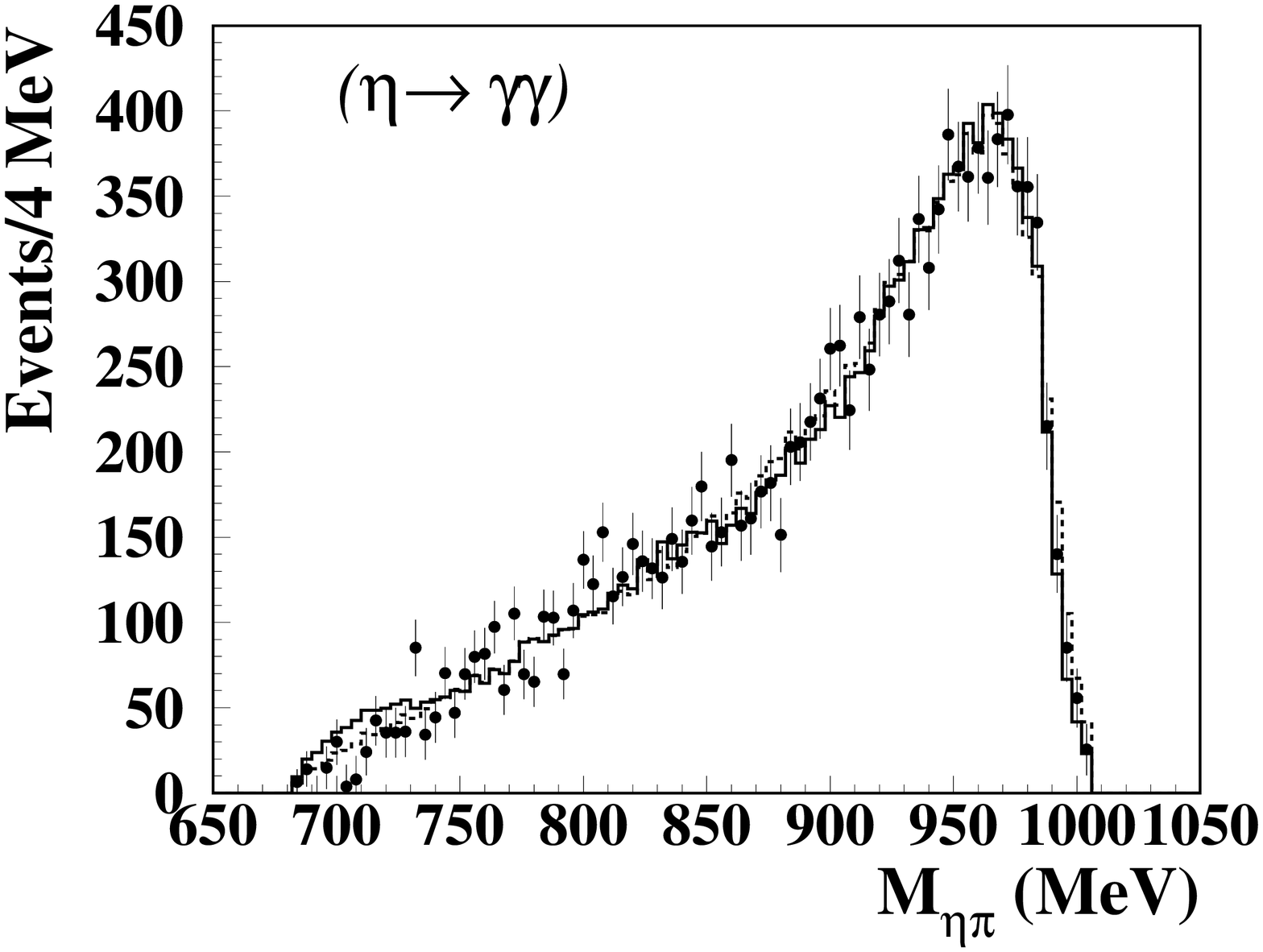}  &
  \includegraphics[width=0.45\textwidth]{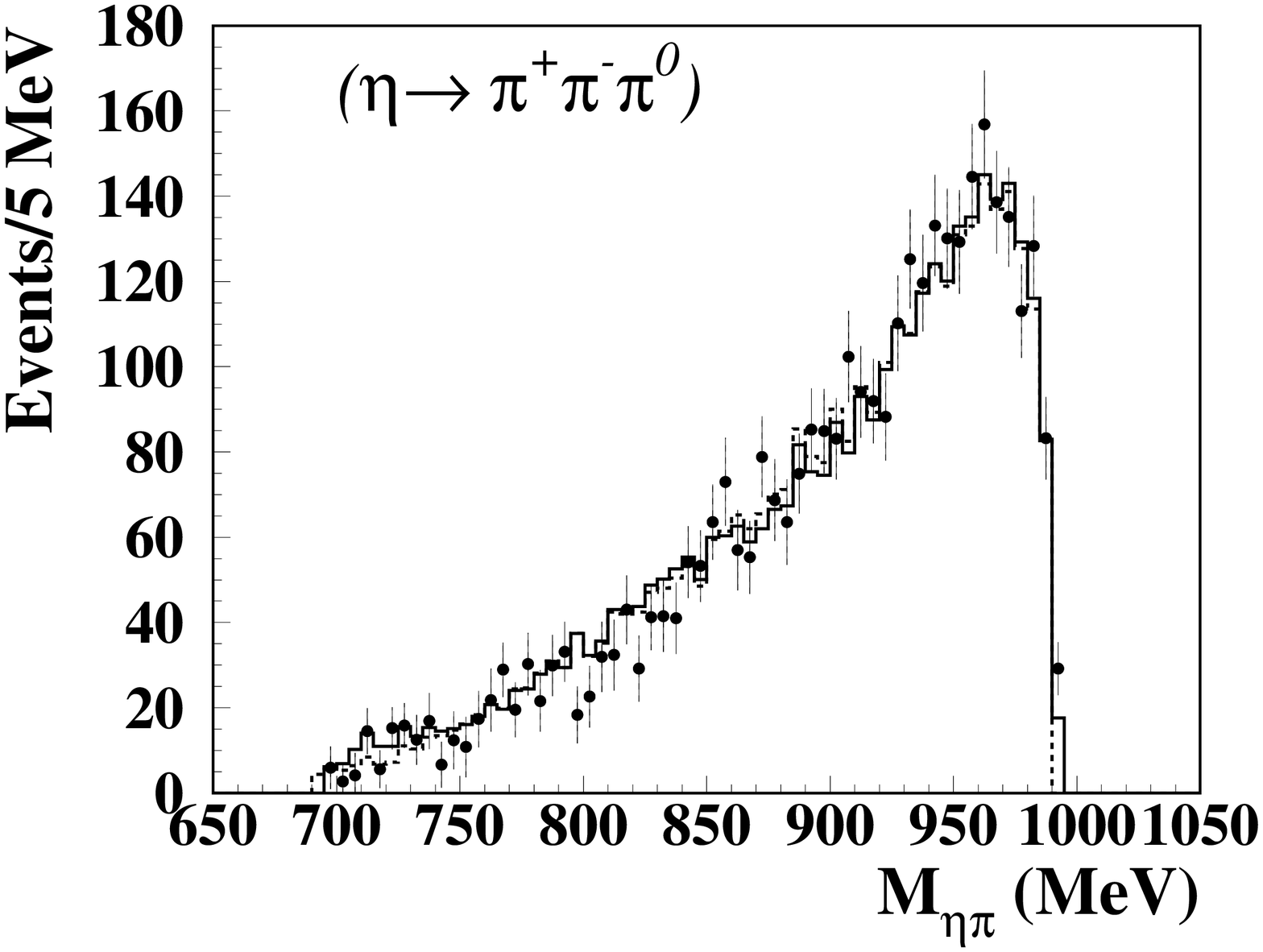}  \\
\end{tabular}
  \caption{ Fit results: points are data after background
  subtraction; histograms represent the fit functions for KL (solid) and NS
  (dashed) models.}
\label{fig:fitplotkl}
\end{figure}
\begin{table}[htb]
  \caption{Correlation coefficients among the relevant $a_0$
      parameters.}
\begin{center}
  \begin{tabular}{lccc||lccc}\hline
    \multicolumn{4}{l||}{KL model} & \multicolumn{4}{l}{NS model} \\
    \hline
    & $M_{a_0}$ & $g_{a_0K^+K^-}$ & $g_{a_0\eta\pi^0}$ & & $g_{a_0K^+K^-}$
    & $g_{a_0\eta\pi^0}$ & $g_{\phi a_0\gamma}$ \\
    $M_{a_0}$             & 1. & & & $g_{a_0K^+K^-}$ & 1. & & \\
    $g_{a_0K^+K^-}$ & 0.931 & 1. & & $g_{a_0\eta\pi^0}$ & -0.565 & 1. & \\
    $g_{a_0\eta\pi^0}$ & 0.584 & 0.550 & 1. & $g_{\phi a_0\gamma}$ & -0.138 &
    0.657 & 1. \\ 
    \hline
  \end{tabular}
\end{center}
  \label{tab:corrkl}
\end{table}
The fit results are shown in Fig.\ref{fig:fitplotkl}, and the parameter
values are listed in Table \ref{tab:fitkl}. 
Good $\chi^2$ probability is obtained for both models. \\
The ratio $R_{\eta}$ is in good agreement with the PDG value
1.729 $\pm$ 0.028\cite{amsler:2008}, confirming that the two samples are
consistent with each other. \\ 
A vector background smaller than the VDM predictions, $(3 \div 5)\times
10^{-6}$\cite{achasov:2001,bramon:1992}, is found in both fits, 
indicating that the $\phi\to\eta\pi^0\gamma$ process is largely dominated
by $\phi\to a_0\gamma$. \\ 
In the KL case, the $a_0$ mass is in agreement with the PDG value (985.1
$\pm$ 1.3) MeV\cite{amsler:2008}.  
A ratio of the squared coupling constants
$R_{a_0}=g^2_{a_0K^+K^-}/g^2_{a_0\eta\pi^0} = 0.58\pm 0.03\pm 0.03$ can be
derived. 
The $g_{\phi a_0\gamma}$ is not a free parameter of this model, but can be
obtained according to the formula: 
\begin{eqnarray}
g_{\phi a_0\gamma} & =           
  \sqrt{\frac{3}{\alpha}\left(\frac{2M_{\phi}}{M^2_{\phi}-M^2_{a_0}}\right)^3 
  \Gamma_{\phi}Br(\phi\to\eta\pi^0\gamma)} =  
  \nonumber \\
\label{eq:gpag}
  & \\
  & = 1.58\pm 0.10\pm 0.16~{\rm GeV}^{-1}\nonumber  
\end{eqnarray}
The $a_0$ width obtained from eq.(\ref{eq:klwidth}) is
$\Gamma_{a_0}(M_{a_0}) \simeq 105$ MeV.\\  
In Table \ref{tab:corrkl} the correlation coefficients among the $a_0$
parameters are shown.\\ 
The couplings $g_{a_0K^+K^-}$ and $g_{a_0\eta\pi^0}$ of the NS fit and
therefore the ratio $R_{a_0} = 0.67\pm 0.06\pm 0.13$ are in agreement with
the KL values.  
In the NS case $g_{\phi a_0\gamma}$ can be determined directly
and is compatible with the value of eq.(\ref{eq:gpag}).
From this fit a total decay width $\Gamma_{a_0}(M_{a_0}) \simeq 80$ MeV can
be evaluated according to eq.(\ref{eq:nswidth}). \\ 
The systematic uncertainties on the parameters account for: {\it
  (i)} sensitivity to the fixed parameters (the $a_0$ coupling
  to $\eta^{\prime}\pi^0$, $g_{a_0\eta^{\prime}\pi^0}$, and $g_{\phi
  K^+K^-}$ in the KL model, $M_{a_0}$ in the NS model); {\it (ii)}
normalization uncertainty; {\it (iii)} data-MC discrepancy of the fraction
of wrong photon pairings (12\% from data and 14\% from MC). \\ 
\section{Unfolding of the $\eta\pi^0$ invariant mass distribution}
In order to allow a better comparison with other experimental results and
with theoretical models, the invariant mass distribution should be
corrected for resolution and smearing effects. 
Therefore an unfolding procedure has been applied to the $\eta\pi^0$
invariant mass distributions by using the method described in
ref.\cite{dagostini:1995}.  
This is an iterative procedure based on the Bayes theorem, which does not
require the inversion of the smearing matrix. \\
The unfolding has been performed separately on both invariant mass
distributions before the background subtraction. 
The smearing matrices are the same used in the fits described in
Sect.\ref{sec:fit}. \\ 
An initial distribution has to be provided as starting point of the
iterative procedure; the unfolded distributions obtained starting from the
output of the KL fit or from a flat distribution in $M_{\eta\pi^0}$ 
differ by less than 3\%.
This difference has been taken into account in the uncertainty
evaluation. \\  
The bin by bin average of the two unfolded distributions is used to
calculate the differential branching ratio
$(1/\Gamma_{\phi})(d\Gamma(\phi\to\eta\pi^0\gamma)/dM_{\eta\pi^0})$  
reported in Table \ref{tab:unfolding}. 
\begin{table}[htb]
  \caption{Differential branching ratio: $m$ is the bin
      center, the errors are the total uncertainties, and the bin width is
      6.35 MeV.}  
\begin{center}
  \begin{tabular}{cc|cc}\hline
 $m$ & $(1/\Gamma_{\phi})(d\Gamma_{\eta\pi^0\gamma}/dm)\times
    10^7$ & $m$ &
    $(1/\Gamma_{\phi})(d\Gamma_{\eta\pi^0\gamma}/dm)\times 10^7$ \\   
    (MeV)     & (MeV$^{-1}$) & (MeV) & (MeV$^{-1}$) \\
    \hline
    691.53 & 0.06 $\pm$ 0.07 & 850.35 & 2.25 $\pm$ 0.13 \\     
    697.88 & 0.18 $\pm$ 0.10 & 856.71 & 2.35 $\pm$ 0.14 \\
    704.24 & 0.18 $\pm$ 0.12 & 863.06 & 2.27 $\pm$ 0.13 \\ 
    710.59 & 0.31 $\pm$ 0.13 & 869.41 & 2.35 $\pm$ 0.13 \\ 
    716.94 & 0.30 $\pm$ 0.08 & 875.76 & 2.42 $\pm$ 0.16 \\ 
    723.29 & 0.38 $\pm$ 0.11 & 882.12 & 2.59 $\pm$ 0.16 \\ 
    729.65 & 0.53 $\pm$ 0.17 & 888.47 & 2.80 $\pm$ 0.14 \\ 
    736.00 & 0.51 $\pm$ 0.13 & 894.82 & 2.92 $\pm$ 0.19 \\ 
    742.35 & 0.53 $\pm$ 0.05 & 901.18 & 3.18 $\pm$ 0.20 \\ 
    748.71 & 0.67 $\pm$ 0.07 & 907.53 & 3.37 $\pm$ 0.17 \\ 
    755.06 & 0.81 $\pm$ 0.07 & 913.88 & 3.48 $\pm$ 0.17 \\ 
    761.41 & 0.94 $\pm$ 0.10 & 920.24 & 3.67 $\pm$ 0.17 \\  
    767.76 & 0.99 $\pm$ 0.11 & 926.59 & 3.94 $\pm$ 0.17 \\  
    774.12 & 0.99 $\pm$ 0.08 & 932.94 & 4.29 $\pm$ 0.25 \\  
    780.47 & 1.08 $\pm$ 0.09 & 939.29 & 4.63 $\pm$ 0.25 \\  
    786.82 & 1.30 $\pm$ 0.10 & 945.65 & 4.89 $\pm$ 0.21 \\  
    793.18 & 1.27 $\pm$ 0.13 & 952.00 & 5.20 $\pm$ 0.22 \\  
    799.53 & 1.42 $\pm$ 0.28 & 958.35 & 5.40 $\pm$ 0.28 \\  
    805.88 & 1.63 $\pm$ 0.28 & 964.71 & 5.44 $\pm$ 0.33 \\  
    812.24 & 1.71 $\pm$ 0.14 & 971.06 & 5.35 $\pm$ 0.22 \\  
    818.59 & 1.79 $\pm$ 0.16 & 977.41 & 4.94 $\pm$ 0.21 \\  
    824.94 & 1.66 $\pm$ 0.18 & 983.76 & 4.02 $\pm$ 0.19 \\  
    831.29 & 1.82 $\pm$ 0.15 & 990.12 & 2.80 $\pm$ 0.27 \\  
    837.65 & 1.96 $\pm$ 0.12 & 996.47 & 1.51 $\pm$ 0.32 \\  
    844.00 & 2.13 $\pm$ 0.13 &  & \\
    \hline
  \end{tabular}
\end{center}
  \label{tab:unfolding}
\end{table}
The uncertainties are both from statistics (data and MC) and from
systematics. 
The main contribution to the systematic error is the difference between the
two unfolded distributions.
The correlation of the contents of nearest neighbour bins of invariant mass
is about 50\%, for next-nearest neighbour bins is about 20\%, and is
negligible for bin distance greater than two. \\    
An additional uncertainty of 3\% on the absolute scale has to be
considered, according to eq.(\ref{eq:avebr}). \\ 
To check this procedure, the unfolded distribution has been fit to
the KL model, without requiring any smearing matrix. 
The parameters values are in good agreement with those of Table
\ref{tab:fitkl}.
\section{Conclusions}
A high statistics study of the process $\phi\to\eta\pi^0\gamma$ has been
performed, by selecting the decay chains corresponding to
$\eta\to\gamma\gamma$ and $\eta\to\pi^+\pi^-\pi^0$.\\
$Br(\phi\to\eta\pi^0\gamma)=(7.01\pm 0.10 \pm 0.21)\times 10^{-5}$ and
$(7.12\pm0.13\pm0.22)\times 10^{-5}$ respectively have been measured.\\
A simultaneous fit of the two invariant mass distributions has been
performed, which shows that the two samples are consistent with each
other.\\ 
Both models used in the fits, the $\phi-$scalar meson coupling 
through the kaon loop (KL model) and the direct coupling (NS model), are
able to reproduce the experimental $\eta\pi^0$ mass distribution. \\
From the fit results that $\phi\to\eta\pi^0\gamma$ decay is dominated by
$\phi\to a_0(980)\gamma$, since the vector contribution is very small,
$Br(e^+ e^-\to VP\to\eta\pi^0\gamma) < 10^{-6}$.\\ 
The fit allows also the extraction of the $a_0(980)$ mass and its couplings
to $\eta\pi^0$, $K^+K^-$, and to the $\phi$ meson.
The mass agrees at one standard deviation level with the PDG value.
The two sets of couplings obtained from the fits agree with each other.
Using these couplings, a total decay width of the $a_0(980)$ in the range $80
\div 105$ MeV is estimated. 
The ratio $R_{a_0} = g^2_{a_0K^+K^-}/$ $g^2_{a_0\eta\pi^0} \simeq 0.6 - 0.7$
is obtained. 
A large $g_{\phi a_0\gamma}$ has been found ($1.6 \div 1.8$ GeV$^{-1}$)
suggesting a sizeable strange quark content of the $a_0(980)$. 
\section{Acknowledgments}
We thank the DA$\Phi$NE team for their efforts in maintaining low background
running conditions and their collaboration during all data-taking. 
We want to thank our technical staff: G. F. Fortugno and F. Sborzacchi for
their dedicated work to ensure an efficient operation of the KLOE Computing
Center; M. Anelli for his continuous support to the gas system and the
safety of the detector; A. Balla, M. Gatta, G. Corradi, and G. Papalino for
the maintenance of the electronics; M. Santoni, G. Paoluzzi, and
R. Rosellini for the general support to the detector; C. Piscitelli for his
help during major maintenance periods. 
This work was supported in part by EURODAPHNE, contract FMRX-CT98-0169; 
by the German Federal Ministry of Education and Research (BMBF) contract
06-KA-957; by the German Research Foundation (DFG), 'Emmy Noether
Programme', contracts DE839/1-4; by INTAS, contracts 96-624, 99-37; and by
the EU Integrated Infrastructure Initiative HadronPhysics Project under
contract number RII3-CT-2004-506078.
\appendix
\section{Appendix: main formulas of the KL model\cite{achasov:1989}}
\label{app:kl}
The scalar term of eq.(\ref{eq:klformula}) has the form: 
\begin{displaymath}
  \frac{d\Gamma_{scal}}{dm}=
  \frac{2\vert g_{\phi K^+ K^-}g(m)\vert^2
  p_{\eta\pi^0}(M^2_{\phi}-m^2)}{3(4\pi)^2 M^3_{\phi}}\left|
  \frac{g_{a_0K^+K^-}g_{a_0\eta\pi^0}}{D_{a_0}(m)}\right|^2  
\end{displaymath}  
where 
\begin{displaymath}
  p_{\eta\pi^0} 
  =\frac{\sqrt{[m^2-(M_{\eta}-M_{\pi^0})^2][m^2-(M_{\eta}+M_{\pi^0})^2]}}{2m}
\end{displaymath}
The detailed formulation of the KL function $g(m)$ can be found in
ref.\cite{achasov:1989}. 
$D_{a_0}(m)$ is the inverse propagator of the $a_0$: 
\begin{displaymath}
  D_{a_0}(m)=M_{a_0}^2-m^2+\sum_{ab}[Re\Pi_{ab}(M_{a_0})-\Pi_{ab}(m)]
\end{displaymath}
The sum is extended over all the possible two particle decays of the
$a_0$: $ab=\eta\pi^0$, $K^+K^-$, $K^0\bar K^0$, and $\eta^{\prime}\pi^0$. \\
The $a_0$ width is:
\begin{equation}
  \Gamma_{a_0}(m)=\frac{\sum_{ab}
  Im\Pi_{ab}(m)}{m}=\frac{\sum_{ab}g_{a_0ab}^2\rho_{ab}(m)}{16\pi m} 
  \label{eq:klwidth}
\end{equation}
where:
\begin{displaymath}
  \rho_{ab}(m)
  =\sqrt{\left(1-\frac{(m_a+m_b)^2}{m^2}\right)
  \left(1-\frac{(m_a-m_b)^2}{m^2}\right)}  
\end{displaymath}  
The parameters of the scalar term that are determined by the fit are the
$a_0$ mass and the couplings $g_{a_0K^+K^-}$ and $g_{a_0\eta\pi^0}$.
The $a_0$ to $\eta^{\prime}\pi^0$ coupling is fixed either to
$g_{a_0\eta^{\prime}\pi^0}=-\sqrt{2}~{\rm cos}\varphi_P g_{a_0K^+K^-}$
($qq\bar q\bar q$ hypothesis) or to $g_{a_0\eta^{\prime}\pi^0}=2 ~{\rm
  sin}\varphi_P g_{a_0K^+K^-}$ ($q\bar q$ hypothesis), where $\varphi_P$ is
the pseudoscalar mixing angle (the value $\varphi_P =
39.7^{\circ}$ has been used\cite{didonato:2007}).
Another fixed parameter is the coupling of the $\phi$ to the $K^+ K^-$
pair:
\begin{displaymath}
  g_{\phi K^+ K^-}=\frac{M_{\phi}\sqrt{48\pi Br(\phi\to K^+
  K^-)\Gamma_{\phi}}}{(M_{\phi}^2-4M_K^2)^{3/4}} = 4.49 \pm 0.07
\end{displaymath}
\section{Appendix: main formulas of the NS model\cite{isidori:2006}}
\label{app:ns}
The differential decay width of the NS model is the
following:
\begin{eqnarray}
  \frac{d\Gamma_{th}(\phi\to\eta\pi^0\gamma)}{dm} & =
      & \frac{8\pi\alpha}{3}\frac{p_{\eta\pi^0}(M^2_{\phi}-m^2)^3}
      {M^3_{\phi}} \left|\frac{g_{\phi a_0\gamma}
      g_{a_0\eta\pi^0}}{m^2-M^2_{a_0}+i M_{a_0}\Gamma_{a_0}(m)} + \right.
      \nonumber \\   
  \label{eq:nsfunction}
  & & \\
  & & \left. + \frac{{\rm b}_0}{M^2_{\phi}}+\frac{{\rm
      b}_1}{M^4_{\phi}}(m^2-M^2_{a_0})+A_{vect}\right|^2
      \nonumber  
\end{eqnarray}
The resonance width is mass dependent according to ref.\cite{flatte:1976}:
\begin{equation}
\begin{array}{l}
 \Gamma_{a_0}(m) =\Gamma_{\eta\pi^0}(m)+\Gamma_{K^+K^-}(m)+\Gamma_{K^0\bar
    K^0}(m)  \nonumber \\
  \nonumber \\
    {\rm where:} \;\;\; \Gamma_{\eta\pi^0}(m) =\frac{g^2_{a_0\eta\pi^0}}{8\pi
    m^2}p_{\eta\pi^0} ; \nonumber \\
  \nonumber \\
  \Gamma_{K\bar K}(m) = \frac{g^2_{a_0 K^+K^-}}{16\pi
    m}\sqrt{1-(2M_{K}/m)^2}\;\;\;{\rm
    for}\;\;\;m > 2M_{K} ; 
  \label{eq:nswidth} \\
  \nonumber \\
  \Gamma_{K\bar K}(m) = \frac{i g^2_{a_0 K^+K^-}}{16\pi
    m}\sqrt{(2M_{K}/m)^2-1}\;\;\;{\rm for}\;\;\;m < 2M_{K} \nonumber \\
  \nonumber \\
  ({\rm with} \;\; K\bar K= K^+ K^-,~K^0\bar K^0) \nonumber
\end{array}
\end{equation}
The scalar background is parametrized with a polynomial with two complex
coefficients, b$_0$ and b$_1$. 
The vector background, $A_{vect}$, takes into account all processes 
$e^+e^-\to V\to V^{\prime} P_1$ with $V^{\prime}\to P_2\gamma$ 
($V$,$V^{\prime}= \rho$, $\omega$, $\phi$ and $P_{1,2}=\eta$, $\pi^0$).\\

\end{document}